\documentclass[reprint,amsmath,amssymb,aps,prf,longbibliography]{revtex4-2}
\usepackage[utf8]{inputenc}

\usepackage{graphicx}
\graphicspath{ {figures/} }

\usepackage{natbib}

\usepackage{hyperref}

\usepackage[dvipsnames]{xcolor}

\usepackage{bm}
\usepackage{amsbsy}
\usepackage{amssymb}
\usepackage{amsmath}
\usepackage{mathtools}
\usepackage{isomath}

\usepackage{enumitem}
\usepackage{cleveref}
\usepackage{siunitx}
\usepackage{tikz}
\usepackage[normalem]{ulem}

\renewcommand{\v}[1]{\mathbfit{#1}}      
\newcommand{\uv}[1]{\widehat{\mathbfit{#1}}}      
\newcommand{\vu}[1]{\mathbf{#1}}         

\newcommand{\nx}{\uv{t}}
\newcommand{\ny}{\uv{\mu}}
\newcommand{\nz}{\uv{\nu}}

\def \p {\partial}

\renewcommand{\bar}{\overline}

\newcommand{\pd}[2]{\mathchoice{\frac{\p #1}{\p #2}}{\p #1/\p #2}{\p #1/\p #2}{\p #1/\p #2}}

\renewcommand{\epsilon}{\varepsilon}

\DeclareRobustCommand{\legendline}[1]{%
\raisebox{2pt}{\tikz{\draw[-,#1,solid,line width = 0.9pt](0,0) -- (5mm,0);}}%
}

\newcommand{\edit}[1]{{\color{black}#1}}

\begin{document}

\author{Timothy A Westwood}
 \email{t.westwood16@imperial.ac.uk}
\author{Eric E Keaveny}
 \email{e.keaveny@imperial.ac.uk}
\affiliation{Department of Mathematics, Imperial College London, South Kensington Campus, London, SW7 2AZ, UK}

\title{Coordinated motion of active filaments on spherical surfaces}

\begin{abstract}
Coordinated cilia are used throughout the natural world for micronscale fluid transport.  They are often modelled with regular filament arrays on fixed, planar surfaces.  Here, we simulate hundreds of interacting active filaments on spherical surfaces, where defects in the cilia displacement field must be present.   We see synchronised beating towards or about two defects for spheres held fixed.  Defects alter filament beating which causes the sphere to move once released.  This motion feeds back to the filaments resulting in a whirling state with metachronal behaviour along the equator.

\end{abstract}

\maketitle


Motile cilia are slender, flexible and active organelles found throughout the natural world \cite{gibbons1981cilia}.  They provide many swimming microorganisms with a mechanism for propulsion \cite{brennen_fluid_1977} and enable the tissues in larger organisms to pump and move the fluids that surround them.  In our own bodies, cilia are responsible for the proper function of many organs, such as the lungs \cite{sleigh1988propulsion} and brain \cite{faubel2016cilia}, and play critical roles in processes such as fertilisation and embryo development \cite{smith_symmetry-breaking_2019}.  

Fluid motion generated by cilia often relies on the coordinated movement of large groups.  These groups can cover the entire surface of a microorganism, or can be distributed in patches on tissue surfaces.  Cilia-driven fluid flows inspired the seminal works in biological fluid mechanics providing the swimming sheet \cite{taylor_analysis_1951} and squirmer models \cite{lighthill_squirming_1952,blake_spherical_1971}.  Along with experiments utilising cells and model organisms \cite{brumley_hydrodynamic_2012,brumley_metachronal_2015,gilpin2017vortex,pellicciotta2020entrainment,ramirez2020multi} and physical models constructed from colloidal particles  \cite{kotar2010hydrodynamic,bruot2016realizing}, modelling efforts have focused on how the coordinated motion itself emerges as a result of the various mechanisms that are present in these systems.  Both minimal rotor and rower models \cite{golestanian2011hydrodynamic,lenz2006collective,niedermayer2008synchronization,brumley2012hydrodynamic,uchida2012hydrodynamic, hamilton2021changes,lagomarsino2003metachronal,wollin2011metachronal} where the cilia are represented by spherical particles that travel along closed paths, or more detailed, largely computational models \cite{elgeti_emergence_2013,chakrabarti2021multiscale,guirao2007spontaneous,solovev2020global,gueron_cilia_1997,han2018spontaneous,martin2019emergence,osterman2011finding} that treat the cilia as filaments have been able to recover important features of the individual dynamics, such as the spontaneous onset of beating, as well as collective synchronisation and metachronal waves.  

These studies encompass many variations in how the cilia move, but by and large focus on coordination for regularly spaced cilia in one-dimensional chains, or two-dimensional arrays on rigid planar surfaces.  The surfaces of swimming microorganisms, however, are moving through the surrounding fluid and are typically topologically equivalent to a sphere leading to defects in cilia motion that yield intricate flow patterns \cite{gilpin2017vortex}.  Initial studies have shown that rotor chains with paths parallel to the surface form metachronal waves on planar surfaces but instead synchronise on the sphere \cite{nasouri_hydrodynamic_2016}.  Metachronal behaviour is recovered once the rotor path is perpendicular to the surface \cite{mannan_minimal_2020}.  Additionally, recent simulation work \cite{ito_swimming_2019,omori_swimming_2020} with prescribed filament kinematics has shown that cilia motion accounts for the dominant contribution to the viscous dissipation, and hence the details regarding their motion, surface density and distribution can not be ignored when quantifying the propulsive efficiency of ciliates.  In this Letter, we perform simulations that allow filament motion to evolve dynamically and study their coordination on spherical surfaces.  Particular attention is paid to the defects in the filament displacement field due to the spherical topology and whether or not the underlying surface can move subject to the force- and torque-free condition.

We simulate the motion of $M$ active, elastic filaments attached to either a spherical rigid body of radius $R$ or a flat, planar surface.  The system is immersed in a viscous fluid and the inertia of the fluid, filaments, and attached bodies are ignored. Filament motion is driven by a follower-force applied to the filament's free end.  Follower forces have been used previously to mimic dynein motor activity on a filament and the resulting motion has been studied in both two \citep{de_canio_spontaneous_2017,man_multisynchrony_2020} and three \citep{ling_feng_instability-driven_2018,sangani_elastohydrodynamical_2020,fily_buckling_2020} dimensions.

\begin{figure*}[!t]
    \centering
    \includegraphics[width=\linewidth]{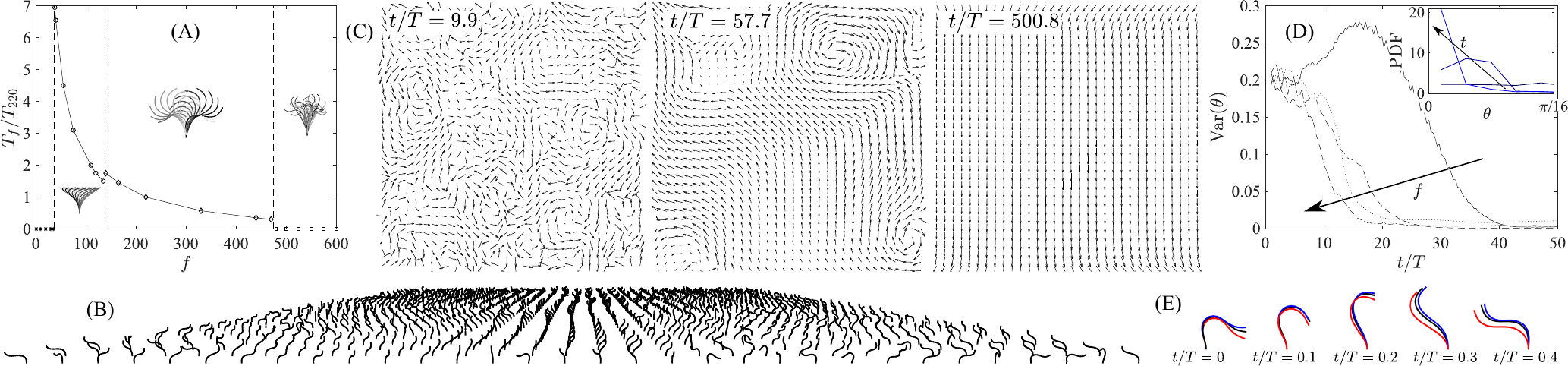}
    \caption{(A): Beat period, $T_f$, and long-time dynamics of an isolated filament (B): Simulation snapshot of $M = 1024$ filaments on a no-slip planar surface. (C): Snapshots of the tip-displacement vector field from the simulation illustrated in (B). (D): The variance of the beat angle, $\theta$, over time for arrays of $M = 400$ filaments for \edit{$f \in \{200, 250, 300, 350\}$}. Inset shows the $\theta$-PDF at different times for $f = 250$. (E): Dynamics of an isolated filament (\legendline{blue}), a filament from the array in (B) (\legendline{black}), and an equatorial filament on a fixed sphere with $R/L = 3$, $\rho = 16.7$ (\legendline{red}) for $f = 220$.}
    \label{fig:fig1}
\end{figure*}
Simulations are performed using the computational framework and models described in \citet{schoeller_methods_2021}.  We summarise the model here and provide a more detailed description in the supplementary material \cite{supp}.  Each filament of length $L$ and cross-sectional radius, $a$, is inextensible, but can bend and twist.  At time $t$, the shape of the filament is described by the space curve $\v{Y}(s,t)$, where $s \in [0, L]$ denotes the arclength.  The orientation of the filament cross-section is described by an orthonormal material frame $\{\nx(s,t), \ny(s,t), \nz(s,t)\}$, where $\nx$ is constrained to be the tangent to the filament centreline through $\nx = \pd{\v{Y}}{s}$. 
Filament dynamics are provided by the force and moment balances, 
\begin{align} \label{eqn:continuous_force_balance}
    \vu{0} &= \p\v{\Lambda}/\p s + \v{f}^H, \\ \vu{0} &= \p\v{M}/\p s + \nx\times\v{\Lambda} + \v{\tau}^H, \label{eqn:continuous_torque_balance}
\end{align}
respectively, where $\v{\Lambda}$ is the internal force on filament cross-sections, $\v{M}$ is the internal elastic moment and $\v{f}^H$ and $\v{\tau}^H$ are, respectively, the forces and torques per unit length exerted on the filament by the surrounding fluid. The internal force $\v{\Lambda}$ arises from the constraint $\nx = \pd{\v{Y}}{s}$ and the elastic moment is expressed in terms of the orthonormal frame vectors as $\v{M}(s,t) = K_B \nx\times\pd{\nx}{s} + K_T\left(\nz\cdot\pd{\ny}{s}\right)\nx,$ where $K_B$ is the bending modulus and $K_T$ is the twist modulus.  At the distal end ($s = L$), the filament is moment-free, $\v{M}(L,t) = \vu{0}$, whilst the force is given by the compressive follower-force
\begin{align}
    \v{\Lambda}(L,t) = -fK_B\nx(L,t)/L^2,
\end{align}
whose magnitude is controlled by the dimensionless parameter $f$.  The force, $\v{\Lambda}(0,t)$, and moment, $\v{M}(0,t)$, at the filament base act to enforce a clamped-end condition.

Each filament is divided into $N$ segments of length $\Delta L = L/N$ and \cref{eqn:continuous_force_balance,eqn:continuous_torque_balance} are discretised to obtain force and torque balances for each of the segments.  This results in a low Reynolds number mobility problem whose solution is the velocities and angular velocities of the segments, and those of the spherical body, if present and free to move.  
The segment mobility matrices are provided by the Rotne--Prager--Yamakawa (RPY) tensors \citep{wajnryb_generalization_2013} for both translations and rotations.  For planar surfaces, we utilise versions of these matrices \cite{swan_simulation_2007} that incorporate the no-slip boundary condition.  For spheres, we discretise their surfaces into $N_\mathrm{RPY}$ RPY particles that are constrained to translate and rotate as a single rigid body following \citep{usabiaga_hydrodynamics_2016}.  If the sphere is fixed, the mobility problem is solved with the condition that the surface velocity is zero. For the free sphere, the force and torque on the sphere set to ensure the total force and torque are zero \cite{supp}.

The segment and the body positions and orientations are updated using the second-order backwards differentiation formula (BDF2), taking advantage also of quaternions and geometric time integration \citep{faltinsen_multistep_2001} to handle rotations.  Applying the BDF2 scheme yields a nonlinear system of differential equations that we solve iteratively using Broyden's method \cite{broyden_class_1965}.

In the simulations, each filament is discretised into $N = 20$ segments of length $\Delta L = 2.2a$.  We consider spheres with radius $R/L = 3$ or $R/L = 5$ and discretise their surfaces using $N_\mathrm{RPY} = 7000$ or $N_\mathrm{RPY} = 16500$ RPY particles, respectively.  This resolves the sphere mobility with error less than 0.1\% as in \citep{ito_swimming_2019}.  \edit{We have also checked the convergence of our surface discretisation to ensure that the no-slip condition is well-resolved for the flows generated by the filaments (see \cite{supp}).}    For spheres, the dimensionless filament number density is given by $\rho = 4\pi L^2 \times M/(4\pi R^2)$.  Finally, the positions of these RPY-particles and the filament segments attached to the sphere are distributed nearly uniformly by first seeding the particle along spirals \citep{saff_distributing_1997} and then evolving them under a repulsive potential to reach their final positions.  

We first revisit the long-time dynamics of an isolated follower-force driven filament that is attached to a planar, no-slip surface.  Varying $f$, we observe (see \cref{fig:fig1}(A)) the same sequence of bifurcations as in \cite{ling_feng_instability-driven_2018} with the static-to-whirling transition occurring at $f \approx 36.5$, the whirling-to-\edit{symmetric} beating at $f \approx 137.5$, and finally the \edit{symmetric} beating-to-writhing at $f \approx 475$.  These are indicated by the sudden changes in the period of filament motion, $T_f$, at these values.  In our studies of collective dynamics, we consider only planar beating ($137.5 \lesssim f \lesssim 474$) and use $T = T_{220}$ as the timescale when presenting our results.
%
%
\begin{figure*}[t!]
    \centering
    \includegraphics[width=\linewidth]{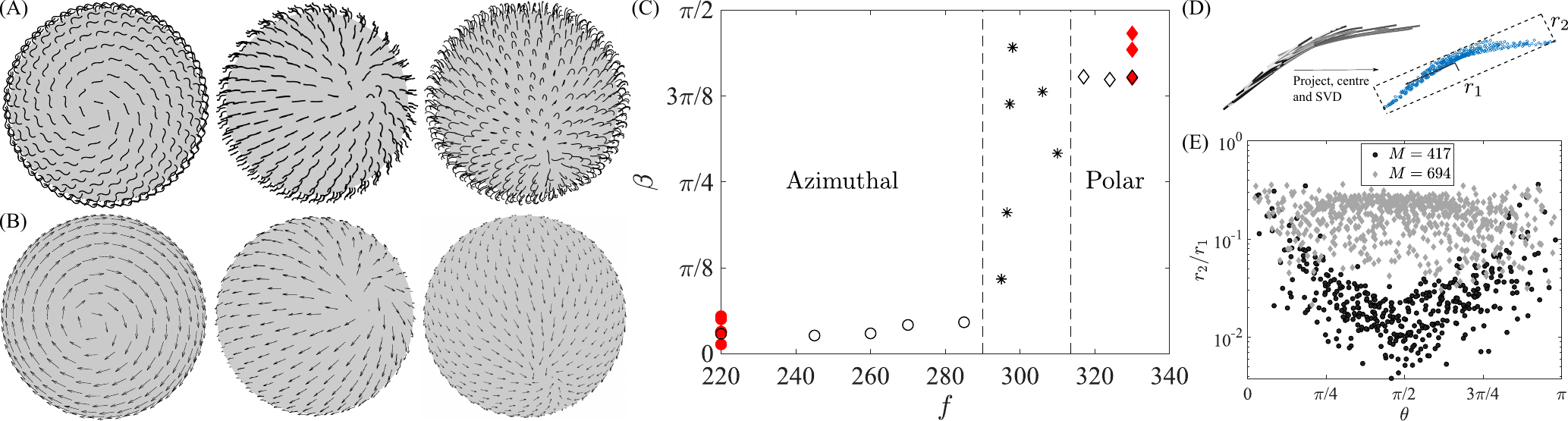}
    \caption{(A): Azimuthal coordination for a fixed sphere with $R/L = 5$, $\rho = 16.7$ and $f = 220$ (left). Polar coordination for fixed spheres with $R/L = 5$, $f = 330$ and $\rho = 16.7$ (middle) and $\rho = 27.8$ (right). (B): Tip-displacement vector fields for the simulations in (A). (C): $f$-dependence of $\beta$ the mean angle between the filament beats and azimuthal directions.  (D): A schematic of the aplanarity measure, $r_2/r_1$, \edit{where $r_1$ and $r_2$ are the dimensions of the rectangle in the tangent plane at the filament attachment point that tightly bounds the filament positions during its beat.}  (E): Variation of $r_2/r_1$ with $\theta$ for $R/L = 5$, $f = 330$ and both $\rho = 16.7$ ($M = 417$) and $\rho = 27.8$ ($M = 694$).}
    \label{fig:fig2}
\end{figure*}

To compare with coordination on spherical surfaces, we examine the collective dynamics for filaments arranged in regular square arrays on planar surfaces.  We find that all filaments eventually synchronise and align their beating.  The synchronisation occurs for all values of $f$ and for all trialed filament spacings ($0.67L - 1.5L$).  A snapshot from one simulation (see video in \citep{supp}) for a $32 \times 32$ ($M = 1024$) array with $f=220$ and filament base separation $1.5L$ is shown in \Cref{fig:fig1}(B).  \Cref{fig:fig1}(C) shows the tip-displacement vector fields at different times taken from this simulation.  The filaments evolve from the random initial conditions, to forming patches of synchronised motion with defects, to alignment and synchrony of their motion.  Filaments near the boundaries do not completely align due to the finite-size of the array.
%
%
The alignment of filament motion is quantified using the distribution of the angle, $\theta$, between the beat plane of each filament and the $x$-direction (see \citep{supp} for computational details).  Filaments at the array boundaries are excluded.  For $M = 400$, the variance of this distribution decreases with time (see \Cref{fig:fig1}(D)) as the distribution itself evolves from being approximately uniform to peaking at $\theta = 0$ as time passes.  We also observe that aligned motion emerges faster at higher $f$ values.  Finally, examining the motion of a representative filament in \Cref{fig:fig1}(E), we see that filament shape is largely unchanged by the collective motion.

Filaments on the surface of a fixed sphere exhibit several differences in their coordination.  First, the defects in tip displacements that emerged, but ultimately disappeared, in the planar case must remain in the spherical one as a consequence of the Poincar\'{e}--Hopf theorem \citep{hopf_vektorfelder_1927}.  Depending on parameter values, we observe both centre and source/sink type defects corresponding to azimuthal and polar beating, respectively.  \edit{At lower $\rho$ values these defects tend to be found close to the defects in the distribution of filament attachment positions, but in general the defect positions appear to depend on the initial conditions or even migrate slowly around the surface.}  As shown in \Cref{fig:fig2}(A) (see also \citep{supp} for videos), simulations with $R/L = 5$, $\rho = 16.7$ ($M = 417$) and $f = 220$ exhibit azimuthal beating while those with $\rho = 16.7$ and $\rho = 27.8$ ($M = 694$), but the higher force value $f = 330$, display polar beating.  For $\rho = 27.8$ and $f = 220$, the simulation did not reach a steady oscillation after $1000T$.  The corresponding tip-displacement vector fields are shown in \cref{fig:fig2}(B).  To investigate further the $f$-dependence of the final state, we compute (see \citep{supp} for details) $\beta$, the average over all filaments of the angle between the final beat direction and the azimuthal direction for $R/L = 3$.  \Cref{fig:fig2}(C) shows $\beta$ transition over a small range of $f$ from values close to zero to values of nearly $\pi/2$, indicating the sudden transition from azimuthal to polar beating.   Incidentally, for planar arrays a similar transition was observed with beating along the lattice directions at lower $f$, but along the diagonals at higher values. The qualitative nature of the collective state appears unaffected by changes in $\rho$ (red symbols in \cref{fig:fig2}(C)).
\begin{figure}[b!]
    \centering
    \includegraphics[width=\linewidth]{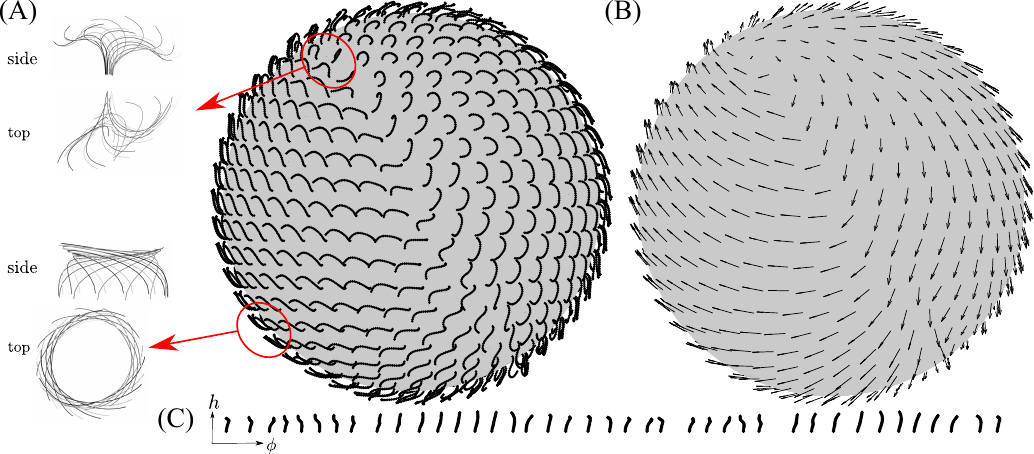}
    \caption{(A): Lab-frame illustrations of the different filament motions on the free $R/L = 5$, $\rho = 27.8$ sphere at $f = 220$, alongside a snapshot of the collective state. (B): The tip-displacement vector field corresponding to (A). (C): An `unwrapped' equatorial band, illustrating the radial tip-displacement $h$ and the azimuthal angle $\phi$.}
    \label{fig:fig4}
\end{figure}

\begin{figure*}[t!]
    \centering
    \includegraphics[width=\linewidth]{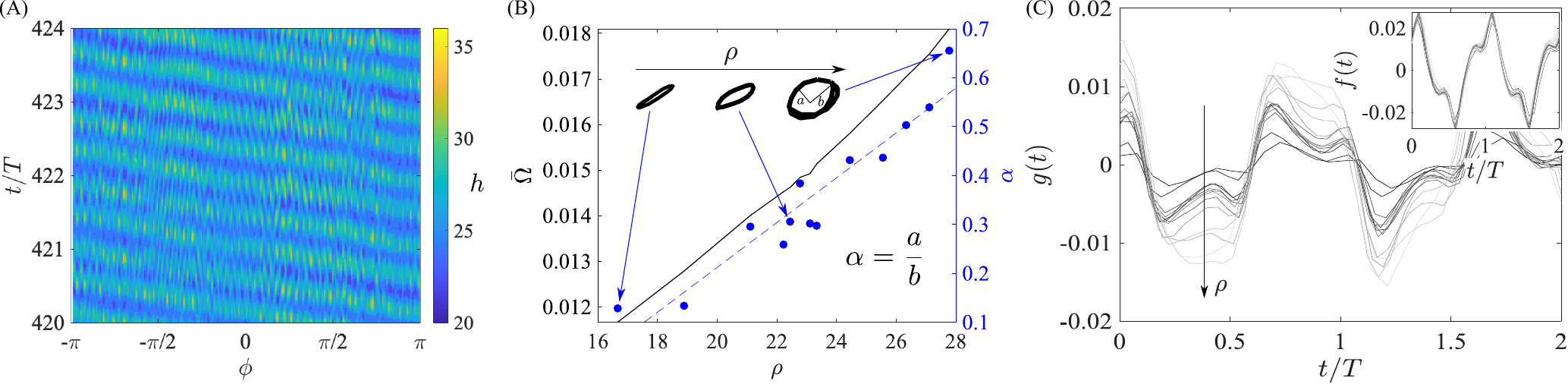}
    \caption{(A): A contour plot of the radial displacement of the filament tip, $h$, against time $t$ and the azimuthal angle $\phi$. (B): The time-averaged angular speed, $\bar{\Omega}$, and $\alpha$, the aspect ratio of the filament base path increase linearly as $\rho$ increases.  (C): Components of $\v{\Omega}(t)$, $f(t)$ and $g(t)$. Lighter curves correspond to higher densities $\rho$.}
    \label{fig:fig5}
\end{figure*}

At low $\rho$ and away from defects, filament motion remains largely unaltered (see \cref{fig:fig1}(E)), however, close to the defects, significant motion out of the expected beat plane is observed.  We measure this deviation using the ratio, $r_2/r_1$, where $r_2$ and $r_1$ are the dimensions of the rectangle (computed using PCA as discussed in \citep{supp}) circumscribing tip displacements over a beat period (see \cref{fig:fig2}(D) for a schematic).  Thus, a planar beat has $r_2/r_1 = 0$, while $r_2/r_1 = 1$ indicates equal maximum displacements in two orthogonal directions.  As shown in \Cref{fig:fig2}(E) for $R/L = 5$, $\rho = 16.7$, and $f = 330$, the out of plane motion increases as $\theta$, the polar angle as measured from the defect position, approaches $\theta = 0$ and $\theta = \pi$ where it has a maximum value of $r_2/r_1 \approx 0.36$.  Increasing to $\rho = 28.7$, altered filament motion is pervasive as aplanarity is observed over the entire surface.

Microorganism and swimming cells that utilise coordinated cilia move freely through the surrounding fluid.  We now explore filament coordination on spheres that are no longer fixed, but move subject to force- and torque-free conditions.  Here, we find that the defect-induced changes in filament motion, especially for high $\rho$, lead to motion of the sphere, which then changes the collective filament dynamics.  For $\rho = 16.7$, collective motion strongly resembles azimuthal beating (recall \cref{fig:fig2}) seen in the fixed sphere case.  The defects coincide with the body's rotation axis \edit{and, as for the fixed spheres, they are often found near the filament placement defects when fixed on the surface, but are also found elsewhere and even found to be mobile in many cases.}  Unlike for fixed spheres, only azimuthal beating occurs for all investigated force values between $f=220$ and $f=330$ on free spheres of radius $R/L = 3$ and $\rho = 16.7$.  Increasing the filament density to $\rho = 27.8$ reveals a novel coordinated state owing to the motion of the underlying surface and was observed for all values of $f$ between $f = 220$ and $f = 330$ on the $R/L = 3$ sphere.  As depicted in \cref{fig:fig4}(A) and (B) (see also \citep{supp} for a video), the majority of filaments divide into two hemispheres: one where they whirl clockwise (when viewed in the body frame), and the other where whirling is anti-clockwise.  While the whirling motion itself is reminiscent of that observed for isolated filaments for $36.5 \leq f \leq 137.5$, the motion here results from the filament base moving due to the sphere wobbling, as shown in \cref{fig:fig4}(A).  In the equatorial region where the hemispheres meet, the individual filaments undergo a flapping-type motion (see \cref{fig:fig4}(A)) and propagate a metachronal wave around the equator.  The position of the rotation axis, \edit{and hence the defects}, coincides with the peak of the metachronal wave \edit{and the defects in the filament attachment positions are found to be in the vicinity of the equatorial region.}  The metachronal behaviour is shown in \cref{fig:fig5}(A) which provides the radial filament tip-displacement, $h$, (see \cref{fig:fig4}(C)) as a function of the azimuthal angle $\phi$ and time $t$.  The diagonal banding is consistent with a wave travelling around the equator with a frequency twice that of the filament motion as there are two peaks in the wave.

While the azimuthal beating and whirling-filament collective dynamics described above seem to be different classes of collective motion, a closer examination shows that they are in fact special cases of a more general behaviour that depends continuously on $\rho$. As illustrated in \cref{fig:fig5}(B), filament bases away from defects move in loops that are slender ellipses for low $\rho$, corresponding motion indistinguishable from azimuthal beating, and approach circles as $\rho$ increases.  This is confirmed quantitatively by computing (see \citep{supp} for details) $\alpha$, the trajectory aspect ratio, which increases monotonically with $\rho$ from values close to $0$ to $0.7$ (\cref{fig:fig5}(B)).  Also shown in \cref{fig:fig5}(B) is the time averaged angular speed of the body, $\bar{\Omega}$, which also increases with $\rho$. We find that to very good accuracy the angular velocity can be expressed as $\v{\Omega}(t) = f(t)\v{p} + g(t)\v{p}_\bot$, where $\v{p}$ and $\v{p}_\bot$ are PCA-computed (see \citep{supp}), orthonormal vectors spanning a body-fixed plane.  The components $f(t)$ and $g(t)$ are shown in \cref{fig:fig5}(C).  While the component $f(t)$ varies only slightly as $\rho$ increases, $g(t)$, which is out of phase with $f(t)$, grows in magnitude.  It is this increase that is responsible for the corresponding increases in both $\alpha$ and $\bar{\Omega}$.   At low $\rho$, $g(t)$ is close to zero and $\v{\Omega}(t)$ is close to an oscillatory rotation about a fixed axis, yielding the slender-loop trajectories of the filament bases (small $\alpha$).  As $\rho$ increases, so too does $|g|$, which widens the filament base paths. 

It is clear that the effects of surface topology and motility conspire to yield large qualitative changes in filament collective dynamics, however, it is interesting to note that in our case the surface motion is purely due to rotations.  We do not observe any significant translation (total displacements are approximately $L$ over $1000T$ of beating), hence swimming, of the body.  Any translation is likely due to slight nonuniformity in the distribution of the filaments over the surface.  For simulations we have performed with very few filaments that are placed at the vertices of Plantonic solids, there is no translation at all, only the wobbling described above.  With this in mind, it is important to recall that the beat of an isolated filament is itself symmetric.  \edit{Polar beating on the fixed sphere retained this symmetry, and releasing the sphere altered the coordinated state rather than yielding symmetry breaking and spontaneous motion, as seen for symmetric phoretic particles at sufficiently high P\'{e}clet numbers \cite{michelin2013spontaneous,michelin2014phoretic}.

Unlike our simulations where the final collective state emerges in the absence of prescribed beat orientations, the collective motion of cilia and flagella used by microorganisms for propulsion are often dictated by different structural elements that restrict beating to particular directions.  A single cilium exhibits an asymmetric beat driven by the action of dynein molecular motors along its length (as opposed to only the tip as in our model) and consists of the well-known effective and recovery strokes \cite{brennen_fluid_1977}.  The beat plane itself is linked with the orientation of the central microtubule pair in the axoneme \cite{satir2014structural}.  With this in mind, it is worth noting that nodal cilia involved in embryogenesis lack this central pair \cite{ishikawa2017axoneme} and are instead observed to whirl \cite{smith_symmetry-breaking_2019}.  Finally, the basal body, and specifically the basal foot, to which the cilia are anchored establishes the beat direction (see for example, \cite{mizuno2017calaxin}) relative to the cell and ultimately, for multicellular organisms, aligns the beat directions with the organism's anterior-posterior axis and fixes the locations of defects in the cilia displacement field.  For example, the somatic cells in Volvox are aligned such that their flagella beat toward defects that coincide with the anterior and posterior poles, ensuring also that the effective stroke is directed toward the posterior \cite{hoops1993flagellar,brumley_metachronal_2015}.  The strong link between beat direction and the colony's anterior-posterior polarity should perhaps be expected.  The anterior-posterior differences in Volvox extend beyond flagella beat alignment, and in particular, the size and number of light-detecting eyespots are greater in the anterior \cite{ueki20105000}.  As flagella provide motility, the correlation between the placement of the sensory organelles used by the colony and its established swimming direction due to flagellar motion are likely needed to enable important responses to external stimuli such as, phototaxis and chemotaxis.  While our simulations do not attempt to capture these important, built-in structural limits on cilia motion and their collective states, they do provide the back drop illustrating that even under minimal restrictions, hydrodynamic interactions between the filaments provides a mechanism for synchronisation, even on spherical topologies.}
%
%

\section*{Acknowledgements}
The authors gratefully acknowledge support from EPSRC Grant EP/P013651/1. TAW is also thankful for funding through an EPSRC Studentship (Ref: 1832024). 
\bibliography{ref_edited.bib,supp.bib}

\end{document}